\documentclass[pre,aps,twocolumn,showpacs,superscriptaddress]{revtex4}
\usepackage{graphicx}
\usepackage{amsthm}
\usepackage{amsmath}
\usepackage{amssymb}
\def\be{\begin{equation}}
\def\ee{\end{equation}}

\def\nw{network}
\begin{document}
\title{Evolving networks with distance preferences}
\author{J. Jost}
\email[e-mail: ]{jjost@mis.mpg.de}
\affiliation{Max Planck Institute for Mathematics in the Sciences\\
        Inselstrasse 22-26, D-04103 Leipzig, Germany}
\affiliation{Santa Fe Institute, 1399 Hyde Park Road, Santa Fe,
New Mexico 87501}
\author{M. P. Joy}
\altaffiliation[Present Address: ]{Children's Hospital, Harvard Medical School, 300 Longwood  Ave., Boston, MA 02115}
\email[e-mail: ]{joy.maliackal@tch.harvard.edu}
\affiliation{Max Planck Institute for Mathematics in the Sciences\\
        Inselstrasse 22-26, D-04103 Leipzig, Germany}
\date{14 June, 2002}

\begin{abstract} We study evolving networks where new nodes when attached
to the network form links with other nodes of preferred distances. A
particular case is where always the shortest distances are selected
(``make friends with the friends of your present friends''). We present
simulation results for network parameters like the first eigenvalue of the
graph Laplacian (synchronizability), clustering coefficients, average
distances, and degree distributions for different distance preferences and
compare with the parameter values for random and scale free networks. We
find that for the shortest distance rule we obtain a power law degree
distribution as in scale free networks, while the other parameters are
significantly different, especially the clustering coefficient.
\end{abstract}


\maketitle

\section{Introduction} Graphs can be considered as substrata of dynamic
networks, and so, several types of graph models have been proposed for
capturing the properties of specific networks \cite{str,bara,doro}. In
particular, evolving networks can be modelled through growing graphs, i.e.
graphs to which continuously new nodes (vertices) and new links (edges)
are added. While regular graphs, i.e. ones where each node has the same
connectivity pattern and where consequently interactions are local in
nature and progress in a slow and orderly fashion from neighbor to
neighbor, can exhibit subtle combinatorial patterns, for a realistic
network model typically a certain amount of irregularity or randomness is
needed. The prototypes here are the random graphs introduced by Erd\"os
and R\'enyi where the connections between the nodes are chosen completely
randomly \cite{bol}. These exhibit quite interesting properties, but often
real networks are not entirely random in this sense, but show some kind of
regularity, not directly in their connectivity pattern, but with respect
to some other variable or order parameter. Such a parameter can be a
clustering coefficient, the average or maximal distance between nodes in
the network (as measured by the minimal number of links separating them),
the distribution of the number of links between the nodes, the correlation
of such properties between neighboring nodes (i.e. those connected by a
link of distance 1), or the first eigenvalue of the graph Laplacian which
is relevant for synchronization properties throughout the network of
dynamic activities at the individual nodes \cite{ws,bara,far,go,jo}.
Models have been proposed that capture some of these aspects. The small
world networks introduced by Watts and Strogatz \cite{ws} are constructed
from regular graphs by creating additional random links between nodes,
with or without deleting some of the existing ones. Once a certain number
of such new links has been introduced in proportion to the number of
regular ones, distances in the graph get dramatically shortened, and,
consequently, activity can spread quite rapidly from a localized source
through the entire network. Another distinct feature of this model is that
there is clustering which is absent in random models. Empirical evidence
is available for the occurrence of clustering in real networks
\cite{new1}. Another interesting model is the one of a scale free network
as introduced by Barab\'asi and Albert \cite{ba,bara}. This is a graph
where new nodes are added and form a fixed number of links with the
existing nodes not completely at random, but with a preference towards
those nodes that already have more connections than other ones. More
precisely, the probability with which  existing nodes receives a link
from a new node is proportional to the number of links it already
possesses. The characteristic feature of the emerging graph here is that
the number of nodes with a given number of links does not decrease
exponentially as a function of the latter as for example in random graphs,
but follows a power law -- the reason why such a graph is called scale
free.  Such models can provide
valuable insights into existing real networks, for example into patterns of
social relations or spreading of diseases in the small world model, or the
connection patterns of internet sites or flight connections between
airports in the scale free model.

It is then a natural question whether there exists an encompassing scheme
that on one hand can put these specific models into a more general
perspective and that on the other hand can offer systematic tools for
analyzing the dependencies among the various network features listed
above. Ideally, these features should depend in an analyzable manner on
certain parameters of the \nw\ construction, and so their
interdependencies could then be studied in terms of relations between the
parameters involved.

We attempt here to take a step in this direction by proposing a general
scheme for constructing evolving \nw s. Our model is characterized by a
distance preference function. This function specifies the probability in
terms of the distance with which an existing node in the network receives
a new link from a newly created node that already has formed one random
link so as to attach it to the \nw\ and to define its distances to the
other nodes. The number of links each node is allowed to make can be
either fixed -- as in our simulation results below -- or also follow some
random distribution. So, for example, we can stipulate that the shortest
distances are always preferred. Thus, a node that is allowed to form a new
link does so preferably to another node of distance 2, i.e. to a direct
neighbor of a node that it is already attached to. This might constitute a
useful model for the formation of social relationships (you want to become
a friend of the friends of your present friends as the easiest or safest
means of forming new relationships). Conversely, we might also stipulate
that always the most distant nodes are the preferred recipients of new
links. Obviously, one then expects that the resulting network has a quite
short average distance between any two nodes, as in the small world and
scale free models. In fact, however, our simulations demonstrate that
directly selecting distances is not as efficient for reducing the
average distance in the network as creating some highly connected
nodes through which many shortest connections can go, as in the scale
free model.  More interestingly perhaps, one may even expect a certain
tendency towards the scale free type when shortest distances are
preferred. Namely, a node that is highly connected then has a greater
chance of receiving a new link than a less well connected one, because the
former has a greater chance of being a direct neighbor of another node
that has received a previous link from a new node that is attaching itself
to the \nw . Thus, we see the principle that the rich get richer that is
characteristic for scale free networks also at work here, although in an
indirect and somewhat mediated form.  A conceptual advantage of this
construction over the scale free one might be that here, for each link, we
only need to evaluate local information, namely check those sites in its
vicinity. More precisely, if we exclusively select sites of distance two
as recipients of new links, then we only have to list all the neighbors of
the present neighbors of the link forming node at each step. In contrast
to this, for the scale free model, the complete connectivity pattern of
any potential recipient anywhere in the \nw\ has to be evaluated. In
general, in our scheme, whether we give preference to short distances or
not, what is crucial for the decision about a new link is not an absolute
property of the candidate as in the scale free model, but rather its
relation, as expressed by the distance, to the link forming node. This may
capture a property that is relevant in some applications.

On the other hand, the scheme where short distances are preferred should
lead to more pronounced local clustering effects and larger average
distances in the network than the scale free construction model. In this
way, we can check that certain \nw\ properties are independent of or at
least not strongly related to each other.

Of course, our scheme also includes the possibility that all distances are
equally preferred. This should generate properties similar to a random
network, although the construction is not entirely identical, because for
a random graph, all nodes are considered equal, whereas here, only those
of the same distance to the node forming links have equal recipient
probabilities, because the distances need not be evenly distributed among
the nodes.

We could also easily supplement our construction scheme by a rule for the
deletion of links and/or nodes according to some criterion to be
specified, as a means to stabilize the size of our \nw . This would allow
a comparison of our model with other ones for evolving networks of given
size range. Here, however, we do not pursue this aspect systematically.

\section{Network construction} We start with a small \nw\ having $m_0$
nodes and then let it grow according to the following scheme. We fix a
number $m$ as the number of connections each new node is allowed to
establish to other nodes existing in the \nw ; in principle, this number
could also be randomly chosen from some distribution instead of being
fixed, but, for simplicity, in our simulations, we only work with a fixed
$m=m_0$, as this will probably not dramatically affect the resulting \nw\
properties. The crucial part of our scheme is the specification of a
probability distribution $p(d)$ for the preferred distance to a node with
which a new link is established. So, when a new node $x_n$ comes in, it is
first allowed to make one connection with a randomly chosen node in the
\nw , in order to attach it to the \nw . (We could also change this rule
and let the first connection prefer well connected recipient nodes, as in
  the scale free model, but in the present paper, we do not perform
  numerical simulations for that rule.) This leaves us with $m-1$ further
links that it is allowed to establish. For the formation of any such link,
we consider a node $x$ in the \nw\ and select it as the recipient of the
new link with a probability given by $p(d(x_n,x))$. Of course, the
formation of any new link changes the distances in the \nw\ and the
creation of further links, until the allotted number $m$ of them has been
formed from $x_n$, then is governed by the new distance pattern. Once
$x_n$ is connected according to this scheme, we create a new node
$x_{n+1}$ and repeat the procedure.

The distance preference function $p(d)$ encodes all the features of our
construction. An important case is where this function is in fact
deterministic, namely where only nodes of distance 2 from $x_n$ are
allowed as link recipients, i.e. the ones that have the smallest possible
distance from it (we are not allowing multiple links, and so no further
link can be attached to a node at distance 1). Another deterministic
choice of $p(d)$ would be to allow only recipients of maximal distance
from $x_n$. This obviously makes the scheme computationally much more
expensive than the exclusive selection of nodes at distance 2.  More
generally, we are interested in distance preference functions $p(d)$ that
are decreasing functions of $d$, i.e. where short distances are preferred
over large ones, but the latter can still be selected with a positive
probability.

 In our simulations as described in the Table~\ref{table}, we consider the
cases where the
  number of links that each new node is allowed to form is $m=2, 3, 4, $
and 5. We let the network grow until its size was 30,000 nodes when we
evaluated the various parameters. We considered three different versions
of the probability for the distances. In Model 1, we exclusively selected
links to nodes of distance 2, i.e. we always formed triangles. In Model 2,
we let the probability be proportional to the inverse distance. Thus,
there was a (slight) preference for shorter distances over larger ones. In
Model 3, in contrast to this, we let the preference function be
proportional to the distance itself (scaled with the maximal distance in
the network). Thus, there is a preference for larger over shorter
distance. Our comparison models are the growing random graph model where
all $m$ links are randomly connected (Model 4) and the scale free or real
world model (Model 5).

In Table~\ref{table} we give the first eigenvalue $\lambda_1$, the
clustering coefficient $C$, the mean path length $L$ and the second moment
of degrees $<k^2>$, for different $m$ values, for Models 1 to 5. The
discussion below will employ the simulation results for $m=5$; as one can
see from the table, the results for $m=3, 4$ are qualitatively similar but
$m=2$ is slightly different. The table gives the averages over 10
simulations each; the standard deviations are quite small.

\begin{table}
\begin{center}
\begin{tabular}{|c|c|c|c|r|}
\hline
$m $& $\lambda_1 $ & $C$ & $L$ & $<k^2>\;  $\\
\hline
 \multicolumn{5}{c}{Model 1}\\
\hline
2  & .00051   & .245980   & 9.9977  &  28.2986  \\
3  & .00089   & .239210   & 7.2686  &  72.4940  \\
4  & .00213   & .219250   & 6.0137  &  140.6150  \\
5  & .00501   & .201360   & 5.2833  &  236.4537  \\
\hline
  \multicolumn{5}{c}{Model 2}\\
%joydist 30k inverse dist jd&&&&\\
\hline
2  & .13906   & .001422  &  7.0212  &  22.3045  \\
3  & .25099   & .001770  &  5.6292  &  48.8695  \\
4  & .32974   & .001981  &  4.9776  &  85.6206  \\
5  & .38889   & .002228  &  4.5795  &  132.6747  \\
\hline
  \multicolumn{5}{c}{Model 3}\\
%joydist 30k dist  jid&&&&\\
\hline
2  & .13872   & .000119  &  7.1207  &  21.7022  \\
3  & .24933   & .000415  &  5.7022  &  47.1782  \\
4  & .32844   & .000681  &  5.0324  &  82.4328  \\
5  & .38688   & .000961  &  4.6203  &  127.5877  \\
\hline
  \multicolumn{5}{c}{Model 4}\\
%randw 30k rnd ra&&&&\\
\hline
2  & .13929  &  .000391  &  7.0690  &  21.9742  \\
3  & .25053  &  .000741  &  5.6659  &  47.9818  \\
4  & .32948  &  .001011  &  5.0061  &  83.8960  \\
5  & .38816  &  .001306  &  4.6017  &  129.8109  \\
\hline
  \multicolumn{5}{c}{Model 5}\\
%realw 30k scalefree rea&&&&\\
\hline
2 &  .15605  &  .000605  &  5.8862  &  39.9532  \\
3 &  .27093  &  .001074  &  4.8676  &  90.2483  \\
4 &  .35066  &  .001482  &  4.3696  &  161.7150  \\
5 &  .40970  &  .001945  &  4.0593  &  250.7354  \\
\hline
\end{tabular}
\end{center}
\caption{The first eigenvalue $\lambda_1$, the clustering coefficient $C$, the mean
path length $L$ and the second moment of degrees $<k^2>$, for Models 1-5, for
different $m$ values.} \label{table}
\end{table}

\section{First eigenvalue}

Spectral properties of small world, scale free and random graph models
have been discussed in \cite{far,go}.  The first (nonzero) eigenvalue of
the graph Laplacian is the crucial parameter for the synchronization
properties of activities at the \nw\ sites as systematically investigated
in our previous work \cite{jo}; see also \cite{rang}. We naturally assume
here that the graph $\Gamma$ under consideration is connected, as are the
graphs resulting from our constructive scheme. Moreover they are symmetric
because we consider undirected links. We label the nodes of $\Gamma$ as
$x_1, x_2,..., x_n$, and we let $k_i$ denote the connectivity, i.e. the
number of neighbors of the node $x_i$. The first eigenvalue is then given
by

 \begin{equation}
   \lambda_1=\inf_{u:\Gamma \rightarrow
\mathbb{R},\ \sum k_{i}u(x_i) =0} \frac{\sum_{x_i \sim x_j} (u(x_i)-u(x_j))^2}{\sum
k_{i} u(x_i)^2}, \label{eq1}
\end{equation}
where $x_i \sim x_j$ denotes that they are neighbors. We can now provide
the following heuristic argument how the creation of a new link in the
\nw\ affects $\lambda_1$ depending on the distance $d(x,y)$ between the
two nodes $x,y$ before the link between them is formed. Namely, for any
function $u$ as evaluated for the infimum in (\ref{eq1}), the new link
only creates an additional summand $(u(x)-u(y))^2$ in the numerator while
the denominator is left unchanged. As the difference in $u$ between
neighbors is minimized for a first eigenfunction, the expected squared
difference $(u(x)-u(y))^2$ should be an increasing function of the
distance between $x$ and $y$. Therefore, the value of a typical candidate
function $u$ for the infimum in (\ref{eq1}) should increase as a result of
the new link in a manner that is positively correlated with the distance
$d(x,y)$. Thus, if our scheme prefers larger distances the first
eigenvalue should get larger than when we select short distances for new
links. Of course, this fits well together with the fact that on one hand,
a larger $\lambda_1$ facilitates synchronization across the \nw , and on
the other hand, connecting nodes that had a large distance should have the
effect of a more pronounced decrease of the average distance which in turn
facilitates synchronization as well.

Our simulations (as described in the Table~\ref{table} ) yield that the
first eigenvalue for Model 1 is .005 which is quite close to the value for
a regular network. Thus, synchronization is quite difficult in such a
network although the average or maximal distance in the network are quite
low (as described below) and the degree distribution of the nodes is quite
similar to the scale free case. In all the other models, $\lambda_1$ is
substantially larger, namely around .39 for Models 2-4 and .41 for Model
5. It might be of some interest that it appears to be about the same or
perhaps even slightly smaller in Model 2, where shorter distances are
preferred, than in the random Model 4, which in turn has a smaller value
than Model 3 with the preference for larger distances. Thus, the scale
free model is the most easily synchronizable of the five, a not always
desirable property.

\section{Clustering} If our distance preference is for the shortest
possible distance, namely 2, then the emerging graph will contain many
triangles, i.e. triples of nodes of mutual distance 1. As a consequence,
we expect that the graph contains highly connected subclusters.

Also, since the creation of any new link increases the first eigenvalue,
it has been suggested by Eckmann and Moses \cite{eck} to employ the number
of triangles for defining some notion of curvature of a graph. This is
based on an analogy with Riemannian geometry where the so-called Ricci
curvature yields a lower bound for the eigenvalue of the Laplace-Beltrami
operator (the Riemannian version of the Laplacian). In other words, the
larger the curvature, the higher the expected value of the first
eigenvalue. As our preceding heuristic analysis of the first eigenvalue of
the graph Laplacian shows, however, there is a problem with the analogy
between the number of triangles and the curvature. Namely, if we add a
link to a given graph, then the expected increase in the eigenvalue is the
higher, the larger the original distance between the two linked nodes was.
In other words, when we select the new link so as to form a new triangle,
the expected eigenvalue increase is smallest, or, when trying to pursue
the analogy with Riemannian geometry, the additional curvature is least.

The clustering coefficient, $C$, of the graph is defined as follows
\cite{new2},
 \begin{equation}
  C = \frac{3 \times \mbox{(number of triangles on the
graph)}}{\mbox{(number of connected triples of vertices)}},
  \end{equation}
where a ``triangle'' is a trio of vertices connected to each other and a ``connected
 triple'' is a vertex connected to an (unordered) pair of other vertices.
 For our choice $m=5$, for a regular network the value for
$C$ is 2/3 (as the number of links of each node is constrained, not all
the neighbors of a given node can be connected among each other, and so
the value is smaller than 1 in any case). In our Model 1, the value 0.20
is quite high, as to be expected, whereas in all other
Models, it is dramatically smaller. In fact, for Model 3 as well as for
the random Model 4, it is even smaller than for the scale free Model 5. In
particular, the difference between the Models 1 and 2 is striking here.

\section{Distances} As already explained, the resulting average or maximal
distance in our \nw\ should be smaller when large distances are preferred
for the establishment of new links. However, this is not so easy to
support through numerical simulations, as in any case, independently of
the preference function adopted, our networks, like the small world and
scale free ones, exhibit rather small maximal distances, say around the
order of 4 or 5 for \nw s with ten or twenty thousand sites, and so the
difference resulting from the preference function cannot be very
pronounced.

There is one observation that can be made here, however. Namely, the
direct preference for forming links to nodes at largest distance is not as
efficient in reducing the average or maximal distance in the network as
the more indirect scheme of preferential attachment to highly connected
nodes employed in the Barab\'asi-Albert model. This demonstrates the
virtue of the latter model. In fact, the average distance $L$ between all
possible pairs of nodes is smallest for that model, namely 4.06, around
4.6 for Models 2-4, and about 5.2 for Model 1. Not surprisingly, a
preference for short connections leads to a larger average distance
although the effect is by no means as pronounced as one might naively
expect. It is surprising, however, that $L$ is slightly larger for Model 3
where large distances are preferred than for the random Model 4, and
slightly smaller for Model 2 with its preference for shorter distances.

\section{Degree distribution} 
\begin{figure}
\includegraphics[width=6cm]{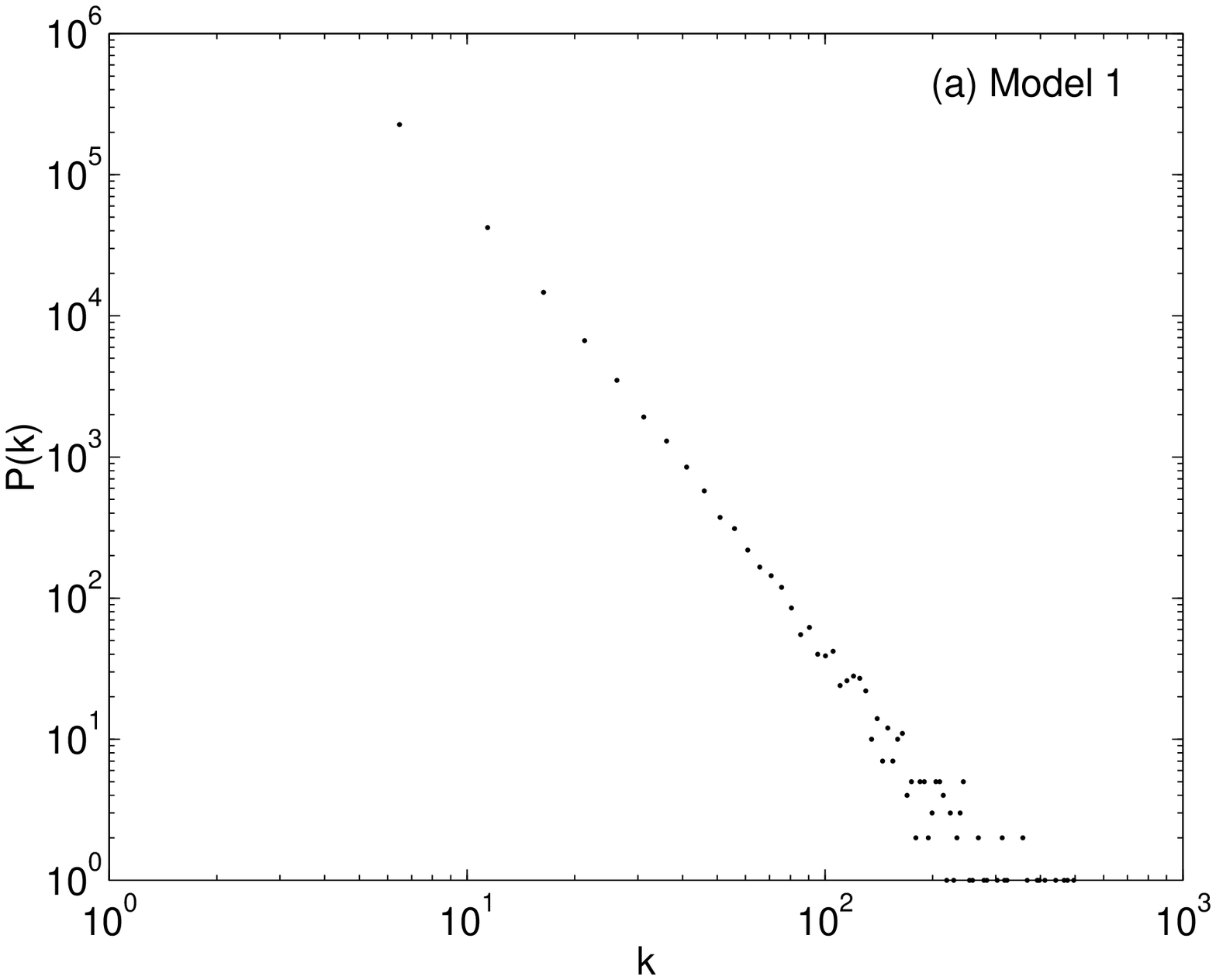}
%\caption{Degree distribution $P(k)$ for Model 1} \label{fig1}
%\end{figure}
%\begin{figure}
\includegraphics[width=6cm]{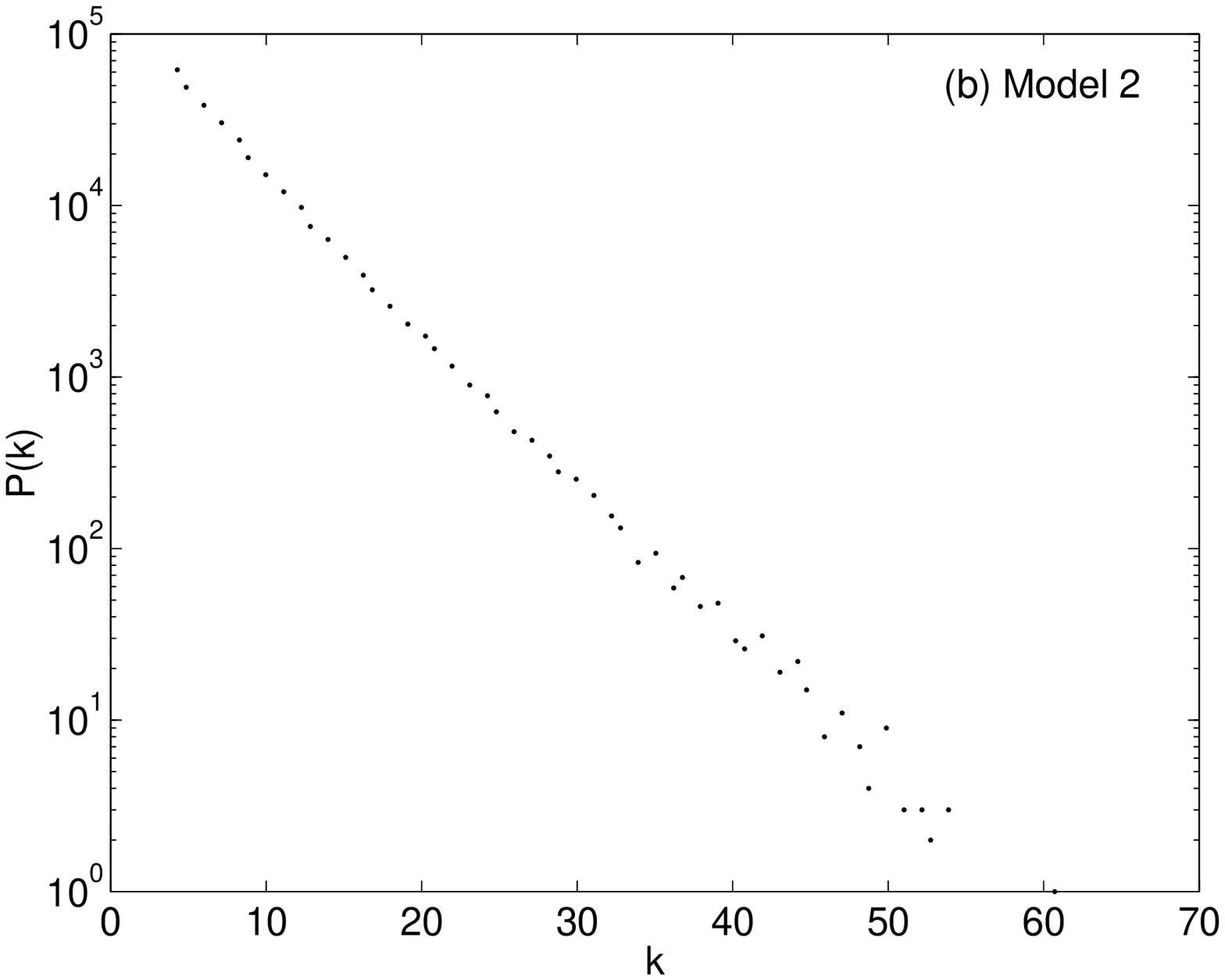}
%\caption{Degree distribution $P(k)$ for Model 2 } \label{fig2}
%\end{figure}
%\begin{figure}
\includegraphics[width=6cm]{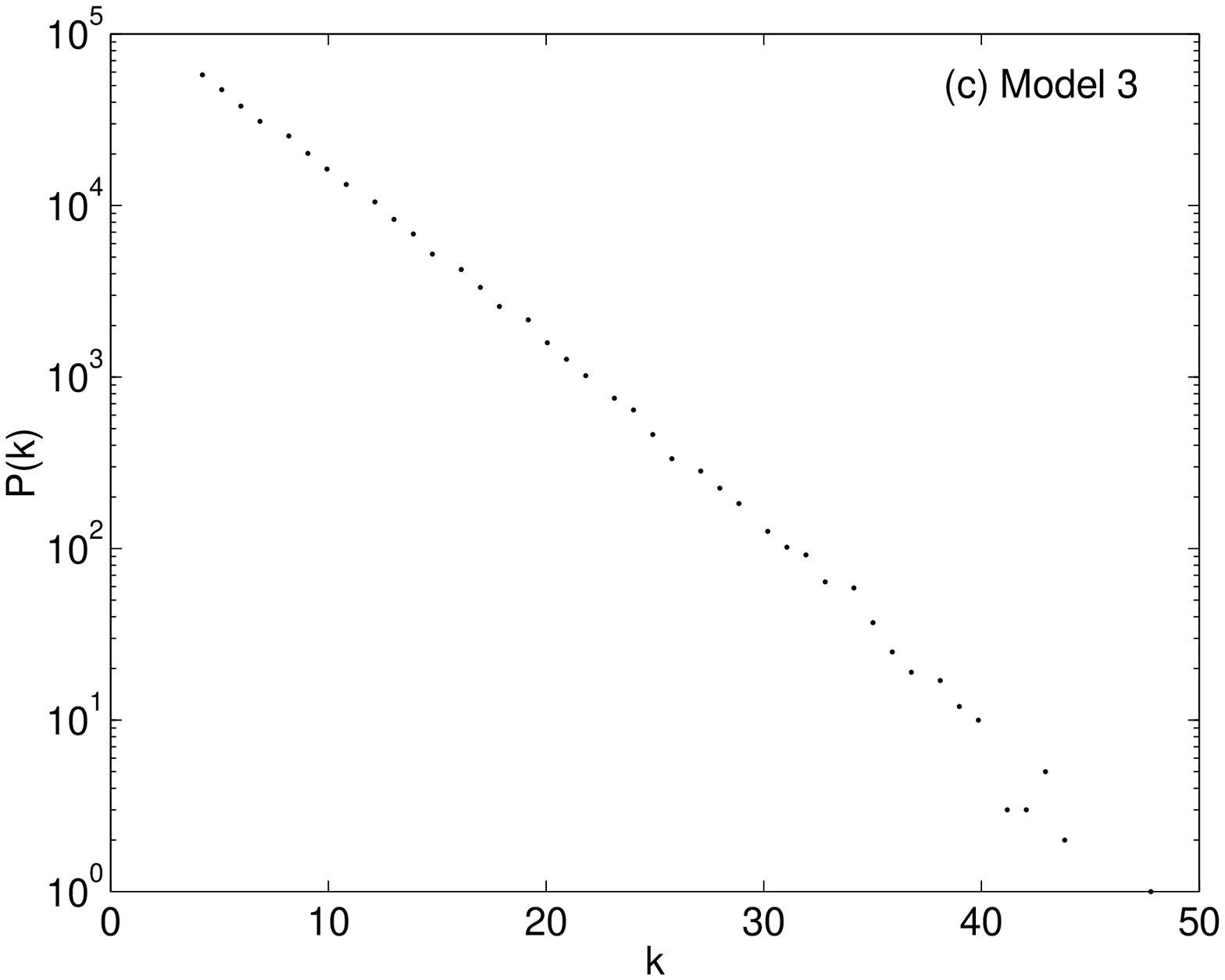}
%\caption{Degree distribution $P(k)$ for Model 3 } \label{fig3}
%\end{figure}

%\begin{figure}
\includegraphics[width=6cm]{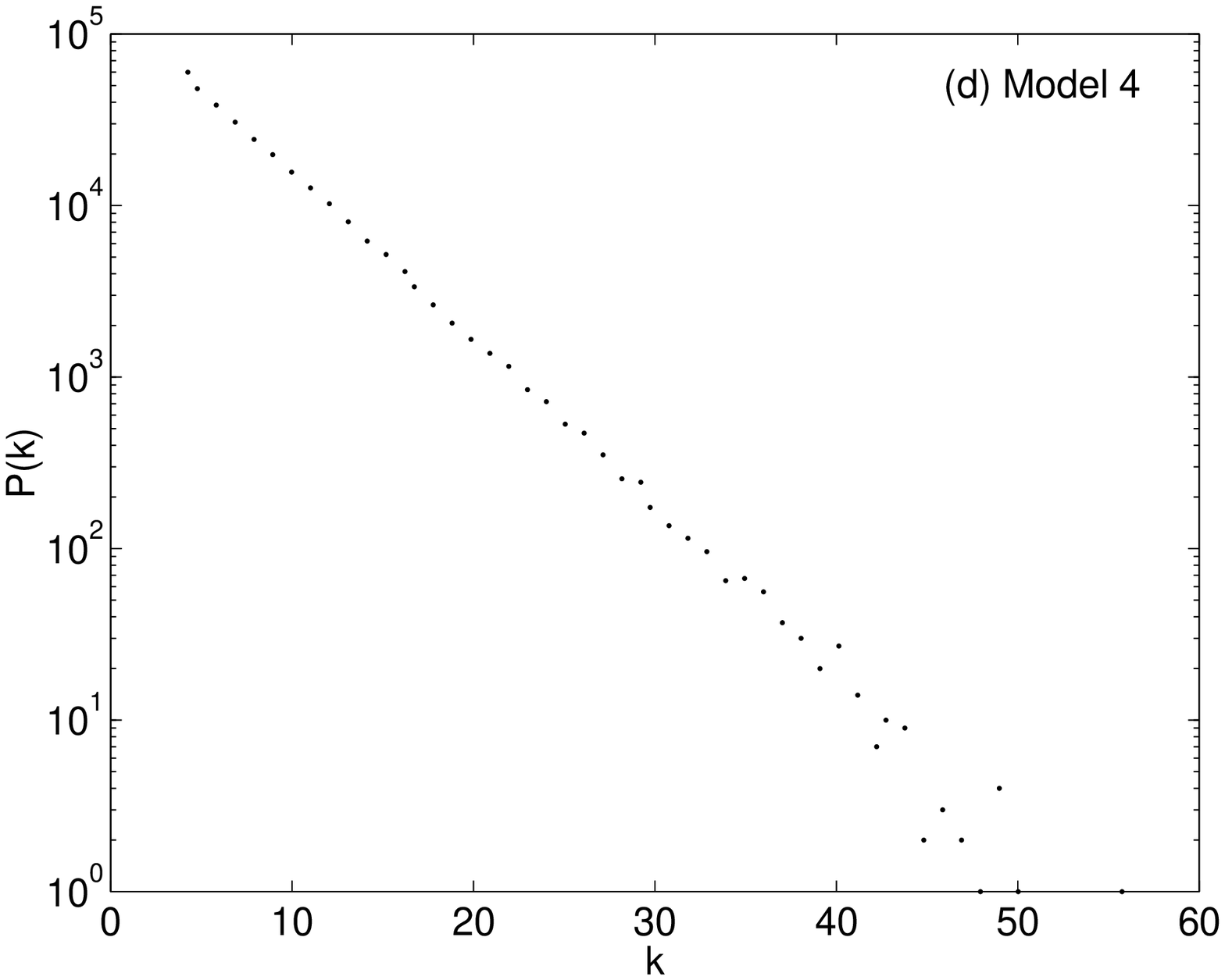}
%\caption{Degree distribution $P(k)$ for Model 4 } \label{fig4}
%\end{figure}
%\begin{figure}
\includegraphics[width=6cm]{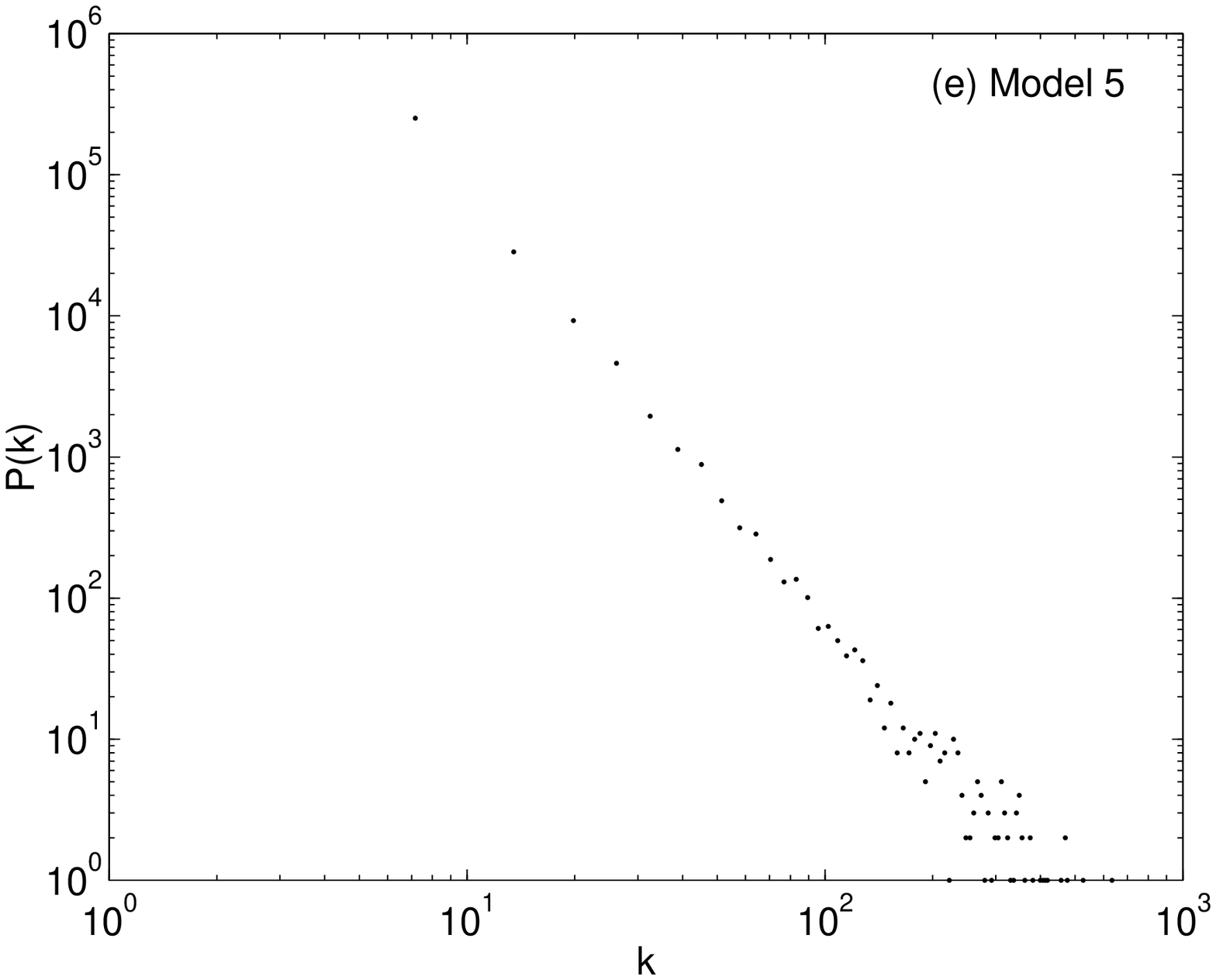}
%\caption{Degree distribution $P(k)$ for Model 5 } \label{fig5}
\caption{Degree distribution $P(k)$ for Models 1,2,3,4, and 5. } \label{fig1}
\end{figure}

One of the distinguishing features of the
scale free or real world model (Model 5) is that the distribution of the
degrees of the nodes decays like a power law in contrast to the
exponential of, for example, the random graph model. In Figure~(\ref{fig1}.a
--\ref{fig1}.e) we give the plots for degree distribution, $P(k)$ for models 1--5,
respectively, with $m=4$. We find that in our Model 1, where exclusively
short connections are selected once a node is anchored in the \nw , the
degree distribution likewise follows a power laws, at least over most of
its regime. (For $m=3$, we get a power law distribution only for some part
of the distribution while the end decays exponentially.) Thus, our
mechanism is capable of producing a network that exhibits a power law
distribution of the degrees but that differs from the scale free model
with respect to a number of distinctive other parameters, like first
eigenvalue and synchronizability, clustering, average distance, etc. In
particular, this feature is independent of those other features.

\begin{figure}
\includegraphics[width=6cm]{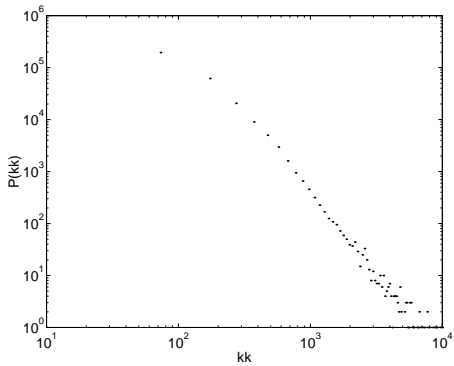}
\caption{Neighbor Degree distribution $P(kk)$ for Model 1 } \label{fig6}
\end{figure}

Models 2 and 3 show an exponential distribution as in the random model
(Model 4). We also find that the distribution of the neighbor degrees
(i.e. the sum of the degrees of all the neighbors of a given node,
$P(kk)$) also partly follows a power law in our simulations for Models 1
and 5. In Fig. \ref{fig6} we plot that for Model 1 with $m=4$.

\section{Correlations} We may ask whether our scheme leads to strong
correlations between neighboring sites in the \nw , with regard to their
connectivity. One possible source of such a correlation in connectivity
could be a correlation in age. Namely, older nodes in the \nw\ have had
more chances than younger ones of receiving a random connection from a new
node, and so, the connectivity should be positively correlated with the
age of a node. However, there is no direct reason why neighboring nodes
should exhibit a pronounced age correlation.

 Another line of reasoning can go as follows: If $x_1$ is a neighbor of a
site $x_2$ of connectivity $l$, then if distance 2 is selected by our
preference function, then $x_2$ has an $l$-fold chance of receiving the
second connection that a new node $x_n$ is making, but the chances of
$x_1$ to benefit from this and receive the third connection that $x_n$ is
making is proportional to $1/l$ as it is facing the competition of the
$l-1$ other neighbors of $x_2$. Thus, the factors cancel, and here, we do
not get an advantage for a node from being a neighbor of a well connected
node. Of course, this heuristic argument does not take the triangle
pattern in the \nw\ into account. We calculated the average of the square
of the degrees of the nodes (second moment), $<k^2>$. The result is given
 in the
last column of the table. The value of this parameter is around 250 for
models 1 and 5 while for models 2, 3 and 4 it is almost half of that
value.

\section{Comparison with other recent \nw\ constructions}

 Dorogovtsev et al. \cite{dms,doro} introduced a model which is similar to
the special case of our Model 1, where each new node forms only two links
and triangles are exclusively selected. They attach new nodes to the
network with links to the two ends of some randomly chosen link already
present in the network.  This scheme depends on the distribution of links
whereas the Model 1 depends on distribution of nodes, though in both cases
triangles are formed.

 Vazquez \cite{va} studied a network where the growth depends on the
knowledge obtained by 'walking' on it. It is a directed graph model unlike
our model. New links are formed with a probability $p$ to a neighbour of a
randomly linked node from the new node and this process is recursively
continued. New nodes are added when there is no new link to form. Beyond a
critical $p$-value it produces scale free network. Here when $p=1$,
neighbours are preferred as in our Model 1 but the process continues
recursively to produce a lot more links of longer distances.

 Jin et al. \cite{jin} introduced a model with fixed number of vertices
where the probability of formation of new links between two nodes depends
preferentially on the number of mutual neighbours. There is a cutoff
on the number of neigbours possible and a possibility for node removal.
This model gives  graphs with high clustering coefficient but there is no
scale free degree distribution.

Holme and Kim \cite{kim} introduced a model that in some respects is
similar to our Model 1. They let the first connection of a new node form
according to preferential attachment as in the scale-free model and then
introduce subsequent links that either form triangles or constitute once
more preferential attachments, according to some random preference. The
resulting \nw\ is again scale free. Their main result is that in a scale
free \nw , the clustering coefficient can take different values (according
to the strength of the triangle preference).

Klemm and Egu\'iluz \cite{kle} consider a growing network model based on
the scale free paradigm, with the distinctive feature that older nodes
become inactive at the same rate that new ones are introduced. This is
interpreted as a finite memory effect, in the sense that older
contributions tend to be forgotten when they are not frequently enough
employed. This results in \nw s that are even more highly clustered than
regular ones.

Davidsen et al. \cite{born} consider a \nw\ that rewires itself through
triangle formation. Nodes together with all their links are randomly
removed and replaced by new ones with one random link. The resulting \nw\
again is highly clustered, has small average distance, and can be tuned
towards a scale free behavior.

\section{Conclusion and discussion} We have introduced a model for
evolving \nw s where each new node, once it is (randomly) anchored to the
\nw , forms further links according to some distance preference function,
and we have compared simulation results for the evolved \nw s with those
for two main types previously considered, namely the random graph model
and the scale free or real world model of Barab\'asi-Albert.
 We found that when always the shortest possible distances are selected
for the recipients of new links, we get a highly clustered \nw\ which is
difficult to synchronize, although it still has a relatively small average
distance between nodes. It also exhibits a power law type behavior for the
distribution of the degrees of the nodes comparable to the scale free
model, although the underlying \nw\ forming mechanism is different, and,
in particular, there is no explicit preference for highly connected nodes
which is considered as the main reason for the power law behavior in the
scale free model.

It has been shown that linear preferential attachment is a necessary
condition for a growing power law network \cite{krap}.  To check this
in our model, we calculated the attachment rate, $\Pi(k)$, as a function
of the degree $k$. To calculate this we used the method described in
\cite{je}.  The attachment rate is numerically fitted with a power law in
$k$ and we obtained the power equal to 1.0 for Model 1 (for $m=5$) and Model 5 and 0.0
for Models 2, 3 and 4. (In Model 1 for smaller values of $m$, this exponent
is less than 1.) This indicates that there is preferential linear
attachment in our Model 1 as in the case of BA model though we don't
explicity introduce that in our model. Surprisingly for Model 2, though it
is similar to Model 1, the attachment rate is independent of the degree as
indicated by the zero exponent of $k$. This explains why the degree
distribution is similar to that of a random one. Even the small probability of 
attaching to second and higher order neighbours in Model 2 produces 
deviation from linear preferential attachment rate. The number of second and 
higher order neighbours are not linearly proportional to the number of first
neighbours of a vertex in these models.

  As the other \nw\ parameters are different from the scale free model,
this shows that this feature is independent of clustering or
synchronizability properties. For other distance preference functions, we
found \nw\ parameters that were roughly comparable with the ones for a
random graph network, and in fact regardless of whether our preference was
proportional or inversely proportional the distance between the link
forming node and the potential recipient.

\begin{acknowledgments} We thank the referee, Mark Newman, and several readers 
of our paper for useful comments.
\end{acknowledgments}

\end{document}